\documentclass[fleqn,usenatbib,useAMS]{mnras}
\usepackage[T1]{fontenc}
\usepackage{ae,aecompl}
\usepackage{graphicx}	
\usepackage{amsmath}	
\title[Viscosity parameter in accretion flows]{Viscosity parameter in dissipative accretion flows with mass outflow around black holes}
\author[S. Nagarkoti and  S. K. Chakrabarti]{
Shreeram Nagarkoti$^{1}$\thanks{E-mail: srnagarkoti@csp.res.in}
and Sandip K. Chakrabarti$^{2,1}$\thanks{E-mail: chakraba@bose.res.in}\\
$^{1}$Indian Centre for Space Physics, Chalantika 43, Garia Station Rd., Kolkata, 700084, India\\
$^{2}$S. N. Bose National Centre for Basic Sciences, JD-Block, Salt Lake, Kolkata, 700098, India
}
\date{Accepted XXX. Received YYY; in original form ZZZ}
\pubyear{2016}
\begin{document}
\label{firstpage}
\pagerange{\pageref{firstpage}--\pageref{lastpage}}
\maketitle
\newpage
\begin{abstract}
Numerical hydrodynamic simulation of inviscid and viscous flows have shown that significant outflows could be
produced from the CENtrifugal pressure supported BOundary Layer or CENBOL of an advective disk.
However, this barrier is weakened in presence of viscosity, more so, if there are explicit energy dissipations
at the boundary layer itself. We study effects of viscosity and energy dissipation theoretically on the outflow rate and show that as 
the viscosity or energy dissipation (or both) rises, the prospect of formation of outflows is greatly reduced, thereby 
verifying results obtained through observations and numerical simulations. Indeed, we find that in a dissipative viscous flow, 
shocks in presence of outflows can be produced only if the Shakura-Sunyaev viscosity parameter 
$\alpha$ is less than $0.2$. This is a direct consequence of modification of the Rankine-Hugoniot relation across the shock 
in a viscous flow, when the energy dissipation and mass loss in the form of outflows
from the post-shock region are included. If we ignore the effects of mass loss altogether, the standing 
dissipative shocks in viscous flows may occur only if $\alpha <0.27$. These limits are tighter than the absolute limit
of $\alpha=0.3$ valid for a situation when the shock itself neither dissipates energy nor any outflow is formed.
We compute typical viscosity parameters required to understand spectral and temporal properties of several black hole candidates such as GX399-4,
MAXI J1659-152 and MAXI J1836-194 and find that required $\alpha$ are indeed well within our prescribed limit.
\end{abstract}
\newpage

\section{Introduction}
Outflows in most of the astrophysical objects are considered to be associated with accretion phenomena \citep{livio97}. 
Indeed, they are generally thought to be produced from the boundary layers of the central objects, including 
black holes (\citealt[][hereafter C96a]{c96a};\citealt{c99,dc99}), paradoxical as it may sound. In a standard Keplerian disk, where the 
centrifugal force is totally balanced by gravity, no boundary layer could be formed around a black hole. This is not the case when 
the angular momentum is `not-Keplerian' such as when the flow is transonic which has low and almost constant angular momentum. 
Here, the outward centrifugal force slows down the infalling matter forming a static or oscillating shock 
(see, C96a and references therein) or, even a shock free flow with a broad and diffused centrifugal 
barrier \citep{c97}. The post-shock region is the so-called CENtrifugal pressure supported BOundary Layer or CENBOL. It has been shown that a large 
region of the parameter space allows such a shock formation \citep[e.g.,][and references therein]{c96b}.  
When the flow has some viscosity, this parameter space starts to shrink \citep{cd04}. 
Furthermore, when the thermal energy at the base of the jet is dissipated, the drive required to form outflows is 
also reduced. So, it is pertinent to ask how the outflow rate depends on both the
energy dissipation and the viscosity at the base of the flow, i.e., the CENBOL. 
Even more important is to know the upper limit of viscosity parameter, when all the 
flow parameters such as the specific energy and angular momentum at the inner edge of the disk are held constant, 
which will still allow the formation of shocks in presence of these dissipations. There are estimates of $\alpha$ parameter in accretion disks
both from observational and numerical simulations. If these estimates are lower as compared to our limits, it would 
indicate that CENBOL would form and radiation emitted from it is an integral part of the spectral and timing properties of black holes.
\defcitealias{nagchak16}{Paper~I}
It has been shown recently (\citet{nagchak16} [\citetalias{nagchak16}]; see also, \citet{kumarchatt13}, with a different viscous stress prescription),  
that there is an upper limit on the \citet{ss73} 
viscosity parameter above which three sonic points and therefore standing shocks are not possible in a viscous accretion flow. These results 
were predicted by \citet{c90a} and \citet{c96a} in the context of isothemal and polytropic flows respectively where it was shown that the 
topology of solutions change dramatically beyond a critical value of viscosity. In Paper I, the limit is shown to be $\alpha_{sup} \sim 0.3$. 
In presence of dissipation at the shock and mass outflow, this limit is likely to change. In the present paper, we address this very important 
issue and show that indeed, shock dissipation tightens the limit to $\alpha_{sup} \sim 0.27$. If, furthermore, outflows are included, then one 
requires to have $\alpha_{sup} \sim 0.2$ for shock formation. This reduces the available parameter space even further.
Nevertheless, as we show below, this limit is not low enough to prohibit formation of
the shocks since even the estimated viscosities in flows from observations are much lower than this value. Thus the observational evidences of shocks 
which decide on the spectral properties of black holes and shock oscillations in explaining quasi-periodic oscillations (QPOs)
remain valid even when outflows from dissipative shocks are present. 

In the next Section, we present the model equations used for our study. In \S 3, we present procedures to solve for
the transonic flows properties, including the shock locations. In \S 4, we discuss how the parameter space,
spanned by the specific energy at the inner sonic point and the specific angular momentum on the black hole horizon 
behave in presence of viscosity, energy dissipation and outflow rate. In \S 5, we show a similar analysis for the data
from outbursts of different black hole candidates. Finally, in \S 6, we draw our conclusions.
\section{Model Equations}

We begin with the governing equations as presented in C96a by considering a stationary, axisymmetric flow in vertical hydrostatic 
equilibrium around a Schwarzschild black hole of mass $M_{BH}$. This is the so-called 1.5 dimensional hybrid model flow
(Chakrabarti, 1989; hereafter C89). The spacetime geometry around the black hole is described 
by pseudo-Newtonian potential first introduced by \citet{pw80}.  
Units of distance and speed are chosen to be $r_g = 2GM_{BH}/c^2$ and $c$, the speed of light respectively.  
Here, $G$ is the universal constant. 
The dimensionless hydrodynamic equations which govern the infall of matter are given by (C96a),

\noindent 
i. The radial momentum equation:
\begin{equation}
v \frac{dv}{dx}+\frac{1}{\rho}\frac{dP}{dx}+\frac{l^{2}_{K}-l^{2}}{x^{3}} = 0,
\end{equation}
ii. Continuity Equation:
\begin{equation}
\frac{d(\Sigma xv)}{dx} = 0,
\end{equation}
iii. Azimuthal Momentum Equation:
\begin{equation}
v \frac{dl}{dx}+\frac{1}{\Sigma x}\frac{d(x^{2}W_{x\phi})}{dx} = 0,
\end{equation}
iv. Entropy Equation:
\begin{equation}\label{ep_eqn}
\Sigma vT\frac{ds}{dx}=\frac{hv}{\gamma -1}\bigg(\frac{dP}{dx}-\gamma \frac{P}{\rho}\frac{d\rho }{dx}\bigg) = Q^+ - Q^-.
\end{equation}

Here, $x$ is the radial distance from the black hole along the disk on the equatorial plane. The local 
variables $v,\ \rho,\ P, \ l_{K}$ and $l$ in above equations are the radial velocity, density, isotropic pressure,
 Keplerian angular momentum and specific angular momentum of the flow respectively.
 Also, $n=\frac{1}{\gamma-1}$, $\gamma$ being the 
polytropic index: $P \propto \rho^\gamma$.
 Define $\Sigma$ and $W$ to be the vertically averaged density and pressure given by $\Sigma=\int_{-h/2}^{h/2}\rho dz=\rho I_{n}h$ 
and $W=\int_{-h/2}^{h/2}Pdz=PI_{n+1}h$ where $I_n=\frac{(2^{n}n)^{2}}{(2n+1)!}$ \citep{matsumoto84}. 
$W_{x\phi}$ is the viscous stress tensor given by $W_{r{\phi}} = -{\alpha}\Pi$ (C96) which is responsible for 
the angular momentum transport where $\Pi=W+\Sigma v^2$. Also, $s$ is the entropy density of the flow, $T$ is the 
local temperature, $Q^+$ and $Q^-$ are the heat gained and lost by the flow (integrated vertically) respectively. 
The thickness of the disk, $h(x)$ is given by the relation $h \sim ax^{1/2}(x-1)$ which is obtained by balancing a component 
of gravitational force and the pressure gradient term in vertical direction, where $a$ is the adiabatic sound speed. 

The heating term, $Q^{+}$, is calculated by using mixed stress prescription (C96a). 
Here, two forms of the viscous shear stresses, $W_{x \phi}(1)=-\alpha \Pi$ and $W_{x \phi}(2)=\eta x\frac{d\Omega}{dx}$ are used,
where, $\eta$ is the dynamic coefficient of viscosity. Heating 
term is calculated using the formula, $Q^{+}=W^{2}_{x\phi}/\eta$, where, $W^{2}_{x\phi}=W_{x \phi}(1)W_{x \phi}(2)$ 
and $\Omega =l/x^2 $. Cooling term $Q^{-}$ is set to zero.
\section{Sonic Point Analysis}
The sonic point analysis is carried out in the standard way (C89, C90a, C96a; Paper I, and 
references therein). After some algebra from the governing equations, we obtain:
\begin{equation}
\frac{dv}{dx}=\frac{N}{D},
\end{equation}
where,
\begin{displaymath}
N = N_{1}-N_{2},~~~~ D= D_{1}+D_{2},
\end{displaymath}
\begin{displaymath}
N_{1}=\bigg[\frac{l^{2}}{x^{3}}-\frac{1}{2(x-1)^{2}}+\frac{a^{2}}{\gamma}\frac{5x-3}{2x(x-1)}\bigg]\bigg[\frac{(\gamma+1)a}{\gamma (\gamma-1)}-\frac{2Aw\alpha ga}{\gamma v^{2}}\bigg],
\end{displaymath}
\begin{displaymath}
N_{2}=\frac{a^{3}}{\gamma ^{2}}\frac{5x-3}{2x(x-1)}+\frac{a}{\gamma}\bigg[\frac{2lAw}{vx^{2}}-\frac{A\alpha w^{2}}{x\gamma v^{2}}\bigg],
\end{displaymath}
\begin{displaymath}
D_{1} = v \bigg[\frac{-(\gamma+1)a}{\gamma(\gamma-1)}-\frac{2Aw\alpha ga}{\gamma v^{2}}\bigg]-\frac{a^{2}}{v \gamma} \bigg[\frac{-2a}{(\gamma-1)}-\frac{2Aw\alpha ga}{\gamma v^{2}}\bigg],
\end{displaymath}
\begin{displaymath}
D_{2}= \frac{Aw\alpha a}{v\gamma} \bigg[1-\frac{ga^{2}}{\gamma v^{2}}\bigg],
\end{displaymath}
\begin{displaymath}
A=-\alpha I_n /\gamma, ~~ g = I_{n+1}/I_n,~\mathrm{and}~~w = ga^{2}+\gamma v^{2}.
\end{displaymath}

At the sonic point, both $N$ and $D$ must become zero simultaneously. Equating $D$ to zero,
we obtain an expression for the Mach number at the sonic point. Likewise, when we equate $N$ to zero, 
expression for sound speed at the sonic points can be calculated. With this, all other variables such as
specific angular momentum and energy of the flow can be calculated. 
Here, inner sonic point $x_{in}$, specific angular momentum at the horizon, namely, $l_{in}$, and $\alpha$ 
are taken as initial parameters to calculate $l$ and $E$ at all radial distances.
The angular momentum and specific energy are calculated using the equations (C96a, Paper I) 

\begin{subequations}
\begin{equation}
l = l_{in} +\frac{\alpha xa^{2}g}{\gamma v}+x\alpha v~~ \mathrm{and}\\
\end{equation}
\begin{equation}
E = \frac{v^2}{2} +na^2 +\frac{l^2}{2x^2}-\frac{1}{2(x-1)}.
\end{equation}
\end{subequations}
\subsection{Procedure to obtain standing shock locations}

A flow must have two `saddle type' sonic points to allow a standing shock formation 
(\citealt[][]{c89}; \citealt{c96a};\citealt[][hereafter C96b]{c96b}).  
The flow originates at the companion surface with negligible radial velocity and gains speed, becoming supersonic after 
passing through the outer sonic point. It then makes a discontinuous jump to the 
subsonic branch at a point where shock conditions are fulfilled and passes through the inner sonic point before 
entering into the black hole supersonically. Several examples of early works are present in C89, C90a and C96a,b. If the 
conditions are not satisfied, the steady state solution remains smooth, though a weaker barrier is produced.

The shock conditions are modified from the usual Rankine-Hugoniot conditions given in standard text books such 
as \citet{ll59}.  To model dissipation from the subsonic post-shock region (CENBOL) which lies between 
the shock and the inner sonic point, we model energy lost between the pre-shock and the post-shok flow as 
$\Delta E=nf(a_{-}^2-a_{+}^2)$, as used by \citet{singhchak12}. 
Also, the mass outflow rate at shock is written as $\dot{M}_o= \dot{M}_- - \dot{M}_+ =\Delta \dot{M}_+$.

Energy, mass and momentum conservation equations at the shock take the form,
\begin{subequations}
\begin{equation}
E_{-}-\Delta E=E_{+},\\
\end{equation}
\begin{equation}
\dot{M}_-=\dot{M}_+ + \Delta \dot{M}_+ = \dot{M}_+ + \dot{M}_o =(1+R_{\dot{M}})\dot{M}_+~~\mathrm{and}\\
\end{equation}
\begin{equation}
\Pi_{-}=\Pi_{+}
\end{equation}
\end{subequations}
where, $R_{\dot{M}}=\frac{\dot{M}_o}{\dot{M}_+}$. In the absence of outflows, $R_{\dot{M}}=0$.
Subscripts `-' and `+' represent pre-shock and post-shock quantities at the shock location, $x_s$. 

Starting from $x_{in}$, we integrate Eq. 5 both inwards and 
outwards to find the flow topology. The subsonic branch for $x>x_{in}$ will be incomplete 
and to determine the outer flow boundary conditions, we assume an arbitrary outer sonic 
point location, say, $x_{out}$ and integrate both inwards and outwards. 
This flow will also pass through our chosen inner sonic point, only when the shock conditions 
are satisfied at heterto unknown $x_s$ located somewhere in between $x_{in}$ and $x_{out}$. We compute shock locations 
following the above procedure, also described in Paper I.
We follow the standard method for calculating the shock conditions (C89, C96a, \citealt{mondal14}). Because of the dissipation and mass outflow taken 
into consideration, the Rankine-Hugoniot conditions at the shock must change. The new, modified relation, which connects the Mach numbers in the pre-shock
and the post-shock flows is given by,
\begin{equation}
\frac{\bigg[\gamma M_{+}^{2}+(\frac{g}{M_+})\bigg]^2}{(1+R_{\dot{M}})^{2}\bigg(n(1+f)+\frac{M_{+}^{2}}{2}\bigg)}
=\frac{\bigg[\gamma M_{-}^{2}+(\frac{g}{M_-})\bigg]^2}{\bigg(n(1+f)+\frac{M_{-}^{2}}{2}\bigg)}.
\end{equation}
\section{Parameter space which allows dissipative shocks}
\subsection{When outflows are absent}

First, we assume $R_{\dot{M}}=0$, i.e., the outflows are absent. For the sake of concreteness, 
we assume a value of $f$ which represents the fraction of available thermal 
energy difference between pre-shock and post-shock regions that is dissipated at the shock, presumably due to inverse Comptonization of soft photons. 
In Eqn. 7a, we assume $f=0.1$ and $0.2$ as examples. Using this method, for each point in the 
parameter space with specific energy at the inner sonic point $\mathcal E$ and the specific angular momentum of flow 
carried to the horizon, $l_{in}$, we found the upper limit of $\alpha$ for which shocks are possible namely, $\alpha_{sup}$.  
In Fig. 1, we present the
parameter space where we plot boundaries of the region in which shocks are present for various values 
of $\alpha$. From right to left, the bounded regions are drawn for $\alpha = 0.001,\ 0.01,\ 0.05,\ 0.1,\ 0.15,\ 0.2,\ 0.25,$ and $\ 0.27$. 
The bounded region practically vanishes for $\alpha > 0.27=\alpha_{sup}$ for $f=0.1$ and at $\alpha > 0.25=\alpha_{sup}$ for $f=0.2$. 
It is to be noted that for the same set of $\mathcal E$ and $l_{in}$ in the parameter space, there could be three sonic points
if $\alpha<\alpha_{sh} <\alpha_{cr} $  (Paper I). If $\alpha <\alpha_{cr}$, not only standing shocks are possible, but also oscillating shocks
are possible as no Rankine-Hugoniot conditions are fulfilled for $\alpha_{sh} < \alpha < \alpha_{cr}$ 
\citep[see, analogous situation in][for inviscid flows]{ryuchakmolt97}. 
In the present paper, we concentrate only on solutions with standing shock waves.  

\begin{figure}
\includegraphics[width=\columnwidth]{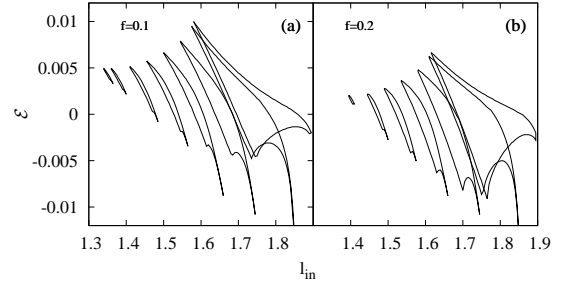} 
\caption{Parameter space which allows dissipative shocks without outflows when (a) $f=0.1$; $\alpha_{sup}=0.27$. (b) $f=0.2$; $\alpha_{sup}=0.25$.}
\label{fig:fig1}
\end{figure}
\subsection{When outflows are present}

In numerical simulations of \citet{mlc94} and later in \citet{mrc96}, it was shown that outflows are 
produced from the post-shock region and matter is ejected in between the centrifugal barrier and the funnel wall of thick advective flow.
These are later identified as the pressure $P=0$ and $\nabla P=0$ surfaces, where $P$ is the gas pressure (C96b). 
  \begin{figure}
\centering
\includegraphics[width=0.6\columnwidth]{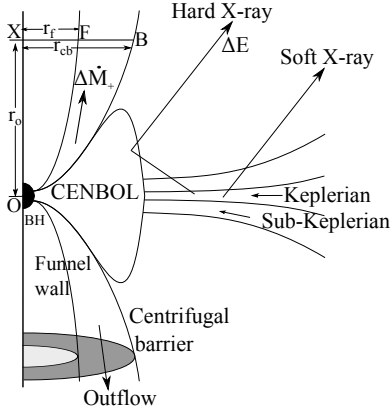}
\caption{Schematic diagram of the centrifugal barrier and funnel wall in which centrifugal pressure driven winds are found to
leave in numerical simulations. Black hole (BH) is at $O$. $OX$ is the axis of rotation. The radius of funnel wall ($r_f$) and the radius of centrifugal barrier ($r_{cb}$) are shown along with the shaded cross-sectional area ($A(r_o)$). Matter lost ($\Delta \dot{M}_+$) at CENBOL pass through this region. Also shown is the energy lost $\Delta E$ through inverse Comptonization, for example. }
\label{fig:fig2}
\end{figure}

In \citet{dc99}, 
this outflow region was used to obtain a complete theoretical solution from the accretion to 
outflow. In \citet{c99} and \citet{das01} 
outflow rates have been computed and it was found that depending on the stength of the standing shock, 
a significant amount of matter may be ejected from CENBOL as outflows. It is to be noted that a
completely new set of one-dimensional equations governing the energy and mass accretion rate conservation
along the z-axis are needed to carry out the outflow properties, since our hybrid model accretion flow
(1.5 Dimensional) is not capable of handling the outflow velocity variation.

In recent numerical simulations which include radiative transfer, it was also found that the CENBOL is responsible 
for the outflows, and the outflow rate starts to decrease with increasing Keplerian disk rate as the CENBOL is cooled 
down by inverse Comptonization \citep{garain12}.  

In what follows, we assume that the CENBOL is the origin of the entire outflow. We assume that the outflow has an angular momentum
same as that at the launching radius. As discussed above, we assume that the boundaries of the outflow are decided by the $P=0$ and $\nabla P=0$ 
surfaces as were observed in numerical experiments mentioned above. This provides us 
with an instantaneous cross-sectional area of the flow which is
required to compute the velocity profile from which the outflow rate is obtained (Das \& Chakrabarti, 1999). 

We consider the outflow to be polytropic in nature and the energy and mass 
conservation equations are written as \citep{dc99}: 
\begin{subequations}
\begin{equation}
E_o=\frac{v_o^2}{2}+na_o^2+\frac{l_o^2}{2r_{m}^{2}(r_o)}-\frac{1}{2(r_o-1)} ~~\mathrm{and}
\end{equation}
\begin{equation}
\dot{M}_o = \rho v_o A_{o}(r_o),
\end{equation}
\end{subequations}
where the variables pertaining to the outflow contain an `o' as the subscript. Here, 
\begin{subequations}
\begin{equation}
r_m(r_o)=\frac{r_{cb}(r_o)+r_f(r_o)}{2} ~~\mathrm{and}\\
\end{equation}
\begin{equation}
A(r_o)=\pi[r_{cb}^2(r_o)-r_{f}^2(r_o)],
\end{equation}
\end{subequations}
where $r_{cb}$, $r_f$ and $A(r_o)$ stand for the radius of the centrifugal barrier ($\nabla P=0$), radius of the funnel wall ($P =0$) 
and the area enclosed between these surfaces respectively, see Fig. 2.
At the centrifugal barrier, the centrifugal force balances gravity \citep{mrc96}.  
At the funnel wall, total effective pressure vanishes. So we can write,
\begin{subequations}
\begin{equation}
\frac{l_{o}^2}{r_{cb}^3}=\frac{r_{cb}}{2r_o(r_o-1)^2}~~ \mathrm{and} \\
\end{equation}
\begin{equation}
-\frac{1}{2(r_o-1)}+\frac{l_{o}^2}{2r_{f}^2(r_o)}=0.
\end{equation}
\end{subequations}
As a result, we get the expressions
\begin{subequations}Fig
\begin{equation}
r_{cb}(r_o)=\bigg[2l_{o}^2r_{o}(r_{o}-1)^2\bigg]^{1/4} ~~\mathrm{and}\\
\end{equation}
\begin{equation}
r_f(r_o)=l_{o}(r_{o}-1)^{1/2}.
\end{equation}
\end{subequations}
Using equations 9(a-b) we calculate locations of sonic points by standard way as explained earlier. We obtain the equation,
\begin{equation}
\frac{dv_o}{dr_o}=\frac{N_o}{D_o},
\end{equation}
where,
\begin{displaymath}
N_o=\frac{a_{o}^2}{A^2(r_o)}\frac{dA(r_o)}{dr_o}+\frac{l_{o}^2}{r_m^3(r_o)}\frac{dr_m}{dr_o}-\frac{1}{2(r_o-1)^2}~\mathrm{and}~~D_o=v_o-\frac{a_o^2}{v_o}.
\end{displaymath}
Following a similar method as earlier, equating $D_o$ with zero, we get the value of Mach no. at the sonic 
point and equating $N_o$ with zero, we get the value of sound speed at the sonic point. Here, we use $E_o=E_+$ and $l_o$ is 
the value of angular momentum of the inflow at the shock location. 

Subsequently, the ratio, $R_{\dot{M}}$ can be calculated using the formula,
\begin{equation}
R_{\dot{M}}=\frac{\dot{M}_{o}}{\dot{M}_+}=\frac{a_{o}^{2n+1}A(r_o)}{2\pi a_+^{2n+1}v_+x_s^{3/2}(x_s-1)}.
\end{equation}
Finally, the ratio of the outflow rate to inflow rate is calculated with the formula,
\begin{equation}
{{{\cal R}_{\dot{M}}}} = \frac{\dot{M}_o}{\dot{M}_-} = \frac{R_{\dot{M}}}{1+R_{\dot{M}}}
\end{equation}

Following the procedure used in the absence of the outflow, we find shock locations as flow parameters are varied.
To be concrete, we chose $f=0.0$, $0.1$, $0.2$ and $0.3$ and obtained the available parameter space having non-zero value of $R_{\dot M}$ 
as shown in Fig. 3 (a-d). The highest values of $R_{\dot M}$ are 0.13,0.1,0.08 and 0.08 when f=0.0, 0.1, 0.2 and 0.3. 
The curves are drawn for $\alpha$ in the same sequence as those in Fig. 1(a). 
From right, $\alpha =0.001,\ 0.01,\ 0.05,\ 0.1,\ 0.15,\ 0.2$, and $\ 0.25$.
It is clear that when outflows are included, the parameter space shrinks significantly as can be seen by comparing Fig. 1(a) and 
Fig. 3(b) when $f=0.1$ was chosen. In Fig. 3(c), results of $f=0.2$ are shown. In Fig. 3(d), for $f=0.3$, the parameter space shrinks even further. 
When $f=0.4$, the parameter space of interest is vanished altogether. This shows that when cooling is 
enhanced, the outflows are choked, supporting the findings of numerical simulations of \citet{garain12}  
that softer states would not have stronger jets (unless there are other effects, such as, magnetic tension which affects the 
CENBOL dynamically and sporadic and powerful, often superluminal, jets are produced). We also find that the 
maximum $\alpha$ called $\alpha_{sup}$ for which shocks are possible is monotonically decreasing. 

\begin{figure}
\centering
\includegraphics[width=\columnwidth]{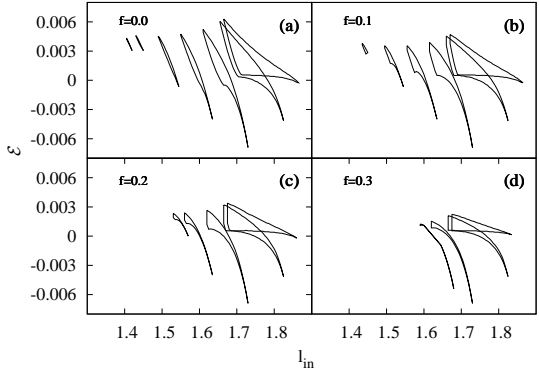}
\caption{Parameter space allowing dissipative shocks in presence of outflows are shown for (a) $f=0.0$; $\alpha_{sup}=0.225$,
(b) for $f=0.1$; $\alpha_{sup}=0.2$,
(c) f=0.2; $\alpha_{sup}=0.125$, (d) f=0.3; $\alpha_{sup}=0.075$.}
\label{fig:fig3}
\end{figure}

It may be instructive to see how the ratio of outflow rate and inflow rate, namely,  ${{\cal R}_{\dot{M}}}$ changes with 
specific angular momentum in the CENBOL. This is because, in numerical simulations of \citet{mlc94}  
it was pointed out that this outflow is primarily centrifugal force driven. If true, ${{\cal R}_{\dot{M}}}$ generally must have a rising trend with 
$l_{in}$. This is also verified from our theoretical work as presented in Fig. 4. As before, the curves are drawn for various $\alpha$
from right to left curve: $\alpha=0.001,\ 0.01,\ 0.025,\ 0.05,\ 0.075,\ 0.1,\ 0.125,\ 0.15,\ 0.175$. 
It is clear from Fig. 4(a) that the outflow rate decreases 
gradually as $\alpha$ increases and finally vanishes when $\alpha>0.175$, when $\mathcal E = 0.0027$. Furthermore, rates are higher
when the $l_{in}$ is also high. From Fig. 4(b), we observe that only stronger 
shocks survive when $\alpha$ is increased. Most interestingly, as already shown for inviscid flows in \citet{c99} and \citet{das01}, 
the highest outflow rate is not necessarily seen when the shock is of highest strength. A balance of the CENBOL size 
and the energy available at the shock plays a major role in deciding this interesting behaviour.  

\begin{figure}
\centering
\includegraphics[width=\columnwidth]{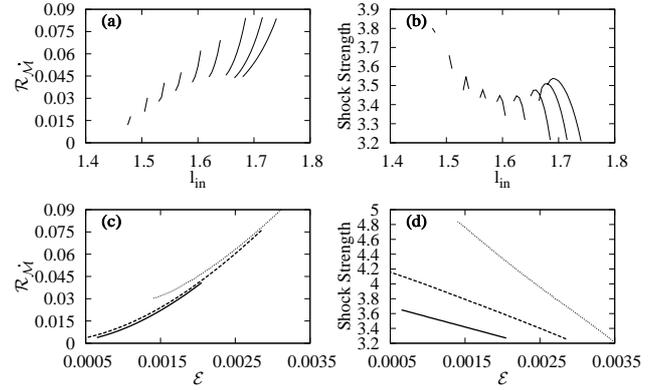} 
\caption{
(a) Variation of ${\cal R}_{\dot{M}}$ at constant energy, $\mathcal E=0.0027$, $f=0.1$ 
for various specific angular momenta with $\alpha=0.001,\ 0.01,\ 0.025,\ 0.05,\ 0.075,\ 0.1,\ 0.125,\ 0.125,\ 0.15,\ 0.175$. 
(b) Behavior of shock strength for the same parameters as (a).
(c) ${\cal R}_{\dot{M}}$ at constant $l_{in} = 1.64$ at $\alpha=0.05$ for various specific energies with
 $f=0.0(dotted),~0.1(dashed)~$and$~0.2(solid)$ lines.
(d) Behavior of shock strength for the same parameters as in (c).
 }
\label{fig:fig4}
\end{figure}

Accordingly, if the base of the jet is more energetic (hot), the outflow rate is expected to be high, even if the 
shock strength is not the highest. In Fig. 4(c-d), we see this behaviour. In Fig. 4(c), we plot ${\cal R}_{\dot{M}}$ as a 
function of $\mathcal E$ for various energy dissipation factor $f$; $l_{in} = 1.64$ throughout. 
The more dissipation the shock has, lesser is the outflow rate and the shock strength. These 
results are very important to link spectral states with outflows. We clearly see that for the 
spectrally softer states (low temperature of the CENBOL; see \citealt[][hereafter CT95]{ct95}) the outflow rate is lower.

Though the incoming transonic flow and the outflow have positive energies, it is not necessary that the energy of 
matter entering into the black hole is positive. This is because the matter may dissipate significant amount of
energy due to cooling processes (by inverse Comptonization, for example), and eventually the flow becomes bound. The outflow
would still be possible with positive energy. This is shown in Fig. 5(a-d) where we demonstrate the general behaviour of  
the shock strength and ${\cal R}_{\dot{M}}$ as functions of specific angular momenta and specific energy.
Viscosity parameter $\alpha=0.05$ was chosen throughout. Each separate curve is for 
different $l_{in}$, changing from $1.6$ to $1.72$ with a step of $0.01$ from right to left. 
For Fig. 5(a and c), there is no dissipation $f=0$ and for Fig. 5(b and d) the dissipation is significant, $f=0.2$. In both the 
cases, the outflow rate generally increases with the specific energy. Lower the energy, higher is the sensitivity. Surprisingly,
the strength of the shock does not have this monotonicity. Rather, it reaches the highest value for an intermediate energy. 

\begin{figure}
\centering
\includegraphics[width=\columnwidth]{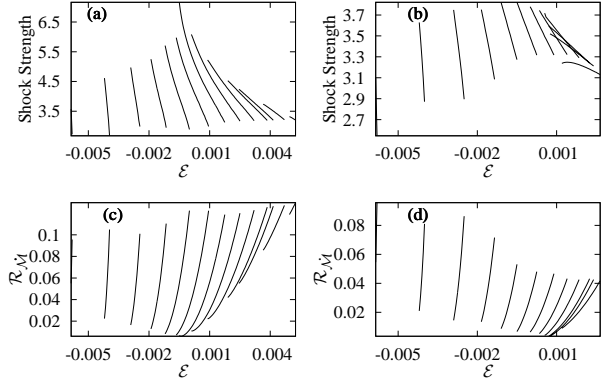} 
\caption{
General behaviour of ${\cal R}_{\dot{M}}$ and shock strength at different cooling in presence of outflow. See text for details.}
\label{fig:fig5}
\end{figure}

\begin{figure}
\centering
\includegraphics[width=\columnwidth]{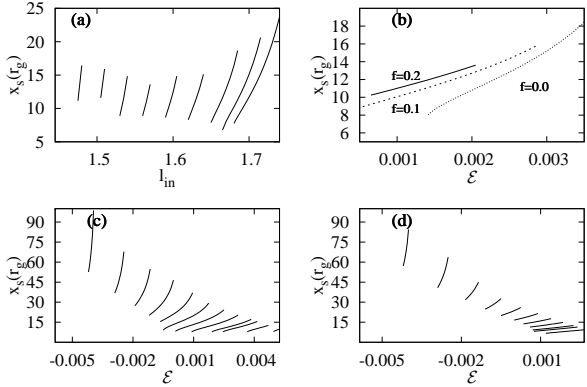} 
\caption{
General behaviour of the shock location against
(a) $l_{in}$ at constant specific energy, $\mathcal E=0.0027$, $f=0.1$ and from right 
to left, $\alpha=0.001,\ 0.01,\ 0.025,\ 0.05,\ 0.075,\ 0.1,\ 0.125,\ 0.15,\ 0.175$,
(b) $\mathcal E$ at constant $l_{in} = 1.64 $, $\alpha=0.1$ with $f=0.2$ (solid), $0.1$ (dashed) and $0.0$ (dotted),
and against $\mathcal E$ at $\alpha=0.05$ at $l_{in}=1.6$ to $1.72$ with a step of $0.01$ from right to left with (c) $f=0.0$ and (d)$f=0.2$.
       }
\label{fig:fig6}
\end{figure}

So far, we have not discussed the actual locations of the CENBOL boundary, i.e., the shock location. Their behaviour is summarized in 
Fig. 6(a-d). In Fig. 6(a), the energy of the flow is kept fixed at $\mathcal E=0.0027$ and the dissipation is also kept at ten percent level
($f=0.1$). Variation of shock location ($x_s$) with different values of $l_{in}$ are drawn 
for $\alpha=0.001,\ 0.01,\ 0.025,\ 0.05,\ 0.075,\ 0.1,\ 0.125,\ 0.15,\ 0.175$ from the right to the left. 
In Fig. 6(b), variation of $x_s$ is shown as a function of $\mathcal E$ when $l_{in}=1.64$ and $\alpha=0.1$ with $f=0.2$ (solid), 
$0.1$ (dashed) and $0.0$ (dotted). The variation of $x_s$ is shown with $\mathcal E$ for $\alpha=0.05$, 
at $l_{in}=1.6$ to $1.72$ with a step of $0.01$ from the right to the left in Fig. 6(c) with $f=0.0$ and in Fig. 6(d) 
with $f=0.2$. It is seen that the shock moves outwards with increasing $l_{in}$ and $\mathcal E$. 
Shock is found to move inwards with increased dissipation.
\section {Estimation of $\alpha$ during outbursts of various sources}

In order to check whether theoretical limits on viscosity parameter make sense in real astrophysical systems,
we choose a well known outbursting black hole candidate named GX339-4. The general
behaviour of the physical parameters have been discussed in \citet{debnath15a}. They analysed the data of 2010-11
outburst of GX339-4 using two component advective flow model of CT95. From their analysis, the time
duration in which the peak of the disk rate followed that of the halo rate was found to be  $\sim 7$ days. 
This we interpret to be the viscous time scale in which the Keplerian matter moves in from the region
where X-rays are insignificant in RXTE energy band. The shock location $x_s$ was found to move 
from $ \sim 150$ to $\sim 20 r_g$ and low frequency QPOs were observed in the range $\sim 0.1$ to $\sim 6 Hz$ in this period. 
\begin{figure}
\centering
\includegraphics[width=\columnwidth]{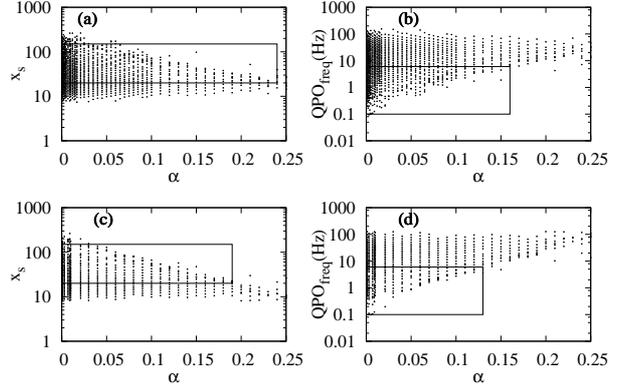} 
\caption{In presence of outflows and in the absence of dissipation at the shock ($f=0$), (a) all possible $x_{s}$, and (b)
QPO frequencies are plotted against $\alpha$. Similarly, when $10\%$ dissipation is considered ($f=0.1$)
in presence of outflows, (c) all possible $x_{s}$, and (d) QPO frequencies plotted against $\alpha$. 
The rectangles contain values relevant for the 2010-11 outburst of GX339-4. }
\label{fig:fig7}
\end{figure}

In Fig. 7, we draw $x_s$ and QPO frequencies against $\alpha$ in presence of outflows as obtained from the 
propagtory oscillating shock (POS) model of Chakrabarti and collaborators \citep{cetal05,cetal08,detal10,detal13,netal12}.
We determine the infall time of matter falling to the black hole using $t_{infall}=\int dt = \int \frac{dx}{v}$ where $v$
is the infall speed obtained from the flow solution for the radial advection of $dx$. The integration is carried out from
the shock location to the inner sonic point. We calculate the frequency of QPO \citep{msc96} as
$ \nu_{QPO}=\frac{1}{t_{QPO}} \approx \frac{1}{t_{infall}}$ in the units of $\frac{r_g}{c}$. 
QPO frequencies are then converted into the units of Hertz by multiplying the value obtained 
from the aforementioned formula by the factor $\frac{r_g}{c}$. We use 
the dynamically obtained mass of GX 339-4, $M_{BH}=7.5 M_{\odot}$ \citep{chen11}.
This formalism has been used previously by \citet{mondal09}. Though we use
the QPO frequency to be equal to the inverse of infall time, as found in \cite{msc96}, this relation need not be 
strictly followed (see also \citealt{d03} and \citealt{sj15}).
 Chakrabarti et al. (2015) pointed out that the resonance condition could
be satisfied over a wide band of parameter range. Thus our computed $\alpha$ would also have a width as determined below. 

Figures 7(a) and (b) are drawn when no thermal dissipation ($f=0$) is taking place at
shock location. Figures 7(c) and (d) are drawn when $10\%$ dissipation is considered ($f=0.1$).
In each Figure, the rectangles contain the region relevant for 2010-11 outburst of GX339-4. The theoretical parameter space
allowing the observed values of QPO frequencies and shock locations in the rectangles indicate 
that allowed $\alpha$ could be as high as $0.13$ from the timing properties.
 Similar analysis of two other outburst sources, namely, MAXI J1659-152 and MAXI J1836-194 have been done. 
Based on the constraints imposed by the spectral and temporal data analysis, we 
find the highest allowed $\alpha$ to be $0.1$ and $0.11$ respectively. These limits of $\alpha$ are 
relevant for the advective component of the accretion disk.  
The uncertainty in $\alpha$ after relaxing the resonance condition would be
$\delta \alpha_{sup}\approx \pm 0.015$, which is insignificant.

\subsection {Analysis using viscous timescale }

In a two component advection flow (TCAF), the viscous Keplerian disk is surrounded by a weakly viscous advective 
component. These two components have two different rates (CT95) which are obtained by fitting
the satellite data with TCAF solution \citep{debnath15a,debnath15b,molla16,jana16,debjit16}. There are ample
observational evidences of these two components as well \citep{smith02}. These components reach close
to the black hole in two different time scales and fitting  of outbursting sources
indicate that the halo component peaks much before the Keplerian component \citep{debnath15a,debnath15b,molla16,jana16,debjit16}.
Difference of these two peak times, assuming that both started at the 
same time near the outer edge would be due to the fact that the Keplerian component is delayed due to viscosity. 
Viscous time is the infall time of the viscous, Keplerian disc component from the outer edge 
to the inner edge of accretion disk. This time can be broken down into the sum of the infall
 time of disk component from outer edge of disk to the shock location and the infall time of matter from shock
 location to the inner edge of the disk. The infall time of the latter component is too small ($\sim$ seconds in our case)
 in comparison to the viscous time ($\sim$ days).
 If $\alpha_{K}>> \alpha_{h}$, where the subscripts $K$ and $h$ are for the Keplerian disc and sub-Keplerian halo components, 
the ratio of the two infall time scales would be $1/\alpha_{K} >>1 $.
Thus the viscous time of the Keplerian component may be safely assumed to be 
the time duration between the peaks of the halo rate and the Keplerian disc rate.
\subsubsection {2010-11 Outburst of GX339-4}

In Fig. $3$ of \citet{debnath15a}
it is shown that there is a time delay of $\sim 7$ days between the peaks of the advective and Keplerian
disk components in the 2010-11 outburst of GX339-4.  This is interpreted to be due to the viscous time delay of the Keplerian component.
In Appendix we show how we calculate the viscous time scale. When we equate our computed time with that of the 
observed delay, we find that $\alpha_{cr} \sim 0.34$ for GX339-4. Thus, during the rising phase of the outburst, 
viscosity parameter has evolved from a smaller value in the quiescent state to at least $\sim 0.34$ when the soft state is reached.
During the peak of the soft state, the value of the viscosity parameter could have gone much higher than this. However, turning the source of 
higher viscosity (e.g., magnetic field entangling inside the disk) off would reduce the matter supply in the Keplerian component and would initiate the declining phase of the outburst. Note that $0.34$ is very much higher that $0.13$, the highest value to form shocks.
Thus the viscosity parameter reaches higher than the critical value in the Keplerian disc component and lower than the critical value in the advective 
component in order to form shocks.
\subsubsection {2010 Outburst of MAXIJ1659-152}

If we follow the same procedure to the outburst of MAXIJ1659-152, we find that though its orbital time is only 2.414 hrs, 
the time duration in which the peak of disk rate followed halo rate was $\sim $ 7 days \citep{debnath15b}.
In the rising state of the outburst, $x_s$ moves from $\sim 360$ to $\sim 40 r_g$  before the soft intermediate 
state is achieved. Low frequency QPOs were observed in the range of $\sim 1.6$ to $6 Hz$. Following an analysis
similar to the previous case, we find that for MAXIJ1659-152, $\alpha \sim 0.22$ would be enough to cause the formation of the soft state. 
Here the advective component can have $0.1$ as the highest value of the viscosity parameter. Thus the TCAF scenario remains consistent. 
\subsubsection {2011 Outburst of MAXIJ1836-194}

\citet{jana16} reported that time duration between the peaks of disk rate and halo rate during 2011 outburst of MAXIJ1836-194
was $\sim 10$ days. Here, the orbital period is 4.8 hrs. 
It was observed that $x_s$ changed in the range of $\sim 32~\mathrm{to}~200 r_g$  in the duration of the outburst.
Low frequency QPOs were observed in the range $\sim 0.4~\mathrm{to}~5.2 Hz$ during the rising phase of 
the outburst. Using the same procedure, we find that the Keplerian component could be formed 
and the soft intermediate state is reached for this source for $\alpha \sim 0.18$. Of course, 
$\alpha$ could be larger when even softer state is reached. In this case also, the advective component viscosity parameter ($0.11$)
is much lower, which is consistent.
\section {Discussion and Conclusions}

It is intriguing that the matter can indeed resist strong gravitational pull of a black hole if its angular momentum does not redistribute 
fast enough to create a Keplerian disc. Under this circumstance, matter forms a centrifugal pressure dominated boundary 
layer from where hot matter may also escape outward driven by the centrifugal force. 
This feature has been explored by several authors \citep{dc99,singhchak12,kumarchatt13}. 
With the rise of viscosity parameter, formation of the centrifugal barrier becomes increasingly difficult. 
One thus requires an answer to the most relevant question in the subject of accretion disc: is viscosity generated inside the
disc sufficient to redistribute low angular momentum matter into a Keplerian disc, 
i.e., maintain a Keplerian distribution in the first place? If not,
then one would have a flow with a centrifugal barrier and a shock so that the hot post-shock region (CENBOL) acts as the Compton cloud 
and produces the observed power-law component as envisaged by CT95. If a Keplerian disk is formed at least on the equatorial 
region due to viscosity higher than a critical value (CT95) its rate would decide if the CENBOL would remain hot. 
Most importantly, if CENBOL forms, so does the outflow. Higher rate in the Keplerian component would cool the CENBOL down and 
produces a spectrally soft state inhibiting the outflow \citep{ct95,c99,garain12}.   

In the present paper, using theoretical approach, we answered many questions. We used the appropriate set of equations governing the 
flow and then modified the Rankine-Hugoniot condition assuming a viscous dissipative flow having both energy and mass loss. 
In Paper I, energy loss and outflows from CENBOL 
were not included. In the present paper we include these after combining the description of the outflow, 
and formalism of dissipation as present in the literature \citep{dc99,singhchak12,kumarchatt13}.  
In \citet{kumarchatt13} $\frac{d\Omega}{dx}$ was chosen to be continuous which triggers a jump in the angular momentum at the shock, 
while we use a continuity in angular momentum since we are discussing axisymmetric shocks. 
This difference induces major changes in the Rankine-Hugoniot conditions, though, some of their conclusions 
remain similar to those present in our work. In \citet{l98} and \citet{l11} 
shocks move outward with increasing $\alpha$ as in our present paper where we kept the conditions on the horizon and 
inner sonic points fixed.

Major conclusions of our work are the followings: 

\begin{enumerate}
\item[] i) Centrifugal pressure supported standing shocks continue to form even when energy dissipation at 
the shock and outflows are included.

\item[] ii) The parameter space shrinks more than what was shown in Paper-I in presence of loss of mass and energy 
at the centrifugal barrier.

\item[] iii) The upper limit of viscosity parameter $\alpha_{sup}$ is reduced to $0.25$ and $0.175$ respectively when only energy dissipation
is included and when both the energy dissipation and outflows are included. This upper limit of $\alpha$ was about $0.3$ when neither
dissipation and nor outflows are included (Paper I). This implies that the parameter space with standing shocks is reduced. 

\item[] iv) The outflow rate increases with lower viscosity and with higher specific angular momentum of the flow. This
directly shows that the outflow is centrifugally driven.

\item[] v) The ratio of the outflow rate to inflow rate need not be maximum for the strongest possible shocks, a result 
which was first presented in Chakrabarti (1999) using a simplified spherical flow geometry.

\item[] vi) As the dissipation at the base of the outflow, namely CENBOL, is increased, the outflow is reduced. This
agrees with detailed numerical simulation results of Garain et al. (2012) that spectrally softer states produce 
lesser outflows.

\item[] vii) Analysis of real data of several black hole candidates show that indeed our theoretical result is very stringent.
In real flow, viscosity parameters required when the shocks are found to be formed appear to be well within our limits. Furthermore,
we compute viscosity parameters in outbursting sources from the time lag of soft X-rays and found that the required viscosity in the
Keplerian component is indeed higher than the critical value. In the advective component, the viscosity is less than the limiting value
required for shock formation. 

\end{enumerate}

In the literature, there are observations and numerical simulations of magnetized disks which show that the 
viscosity parameter in a realistic flow cannot be too high.
The models relying on magneto-rotational instability (MRI) for the transport of angular momentum \citep{bh91,hb92,bra95,hawley95,hawley96}
have produced reliable results in simulating magnetized disks. Such numerical simulations show that 
a value of $\alpha \sim 0.01$ is achieved by such process.
If values of $\alpha$ greater than this are to be achieved, a net vertical magnetic flux condition has to be imposed 
\citep{hawley95,hawley96,sano04,pessah07}. 
Smak (1999) reported that appropriate value of $\alpha$ in the hot, ionized state during dwarf-novae-oubtursts was $\sim 0.1$ in the Keplerian disk. 
But taking the measured time intervals between outbursts into account, \cite{cannizzo88} 
suggested the value of $\alpha$ in the cool state to be in the order of $\sim 0.01$. 
\citet{hk01} indicated $\alpha \sim 0.1$ in the Keplerian component. It is found that most MRI
 simulations find the effective value of $\alpha$  to be of the order 
of $0.01$ while the values calculated from observations are of the order $0.1$ \citep{king07}. 
Since the viscosity of a Keplerian disk has to be above the critical value in order to achieve 
the distribution within the infall time, such a result is expected.
 
Combining these results in conjuction with the constraints on viscosity parameter obtained above,
we find that centrifugal barriers and shocks are allowed in a realistic flow even with 
certain amount of energy dissipation (through inverse Comptonization, for instance) and mass loss (outflows). 
Were the upper limit of $\alpha$ parameter too low, in view of numerical results stated above, the shocks would not 
have been formed and the advective flow component would have just behaved as a dynamic corona without a strong centrifugal barrier. 
However, satisfactory 
fitting of observational data with TCAF \citep{debnath15a,debnath15b,mondal14,jana16,molla16,debjit16}
indicates that the behaviour of CENBOL remained predictable and is indeed an important component of the emission
process. Thus the viscosity in the 
flow cannot be too high \citep{mondal15} to obliterate it, especially in harder states. 
In presence of magnetic fields threading the CENBOL, the outflow rate would be enhanced 
removing some angular momentum. Though there is to date no consistent solution
which suggests the presence of long range poloidal fields in a turbulent disks we consider,
even a small amount of field may affect the dynamics of the flow. So, it would be instructive to 
study the effects of the fields on the general conclusions we draw here. This will be reported in near future.
\section*{Acknowledgments}
Shreeram Nagarkoti acknowledges the support from a fellowship from Abdus Salam International Centre for 
Theoretical Physics, Italy, which allowed him to pursue the present research.
{}
\appendix
\section{Calculating viscous timescale}
The viscous timescale, $t_{visc}$ is given by the relation, $t_{visc} \approx \frac{r^2}{\alpha c_s h} \approx \frac{r^{1/2}(r-1)}{\alpha (\frac{h}{r})^2}$
 where, $\alpha$, $c_s$, $r$ and $h$ stand for the viscosity parameter, the adiabatic sound speed,
 radius at outer edge (in units of $r_*=\frac{Gm}{c^2}$), and the scale height of the accretion disk respectively \citep{p81}.
We use the relation   $\frac{h}{r}= 2.4\times 10^{-3} \alpha^{\frac{-1}{10}}\dot{M}^{\frac{3}{20}}m^{\frac{-3}{8}}r^{\frac{1}{8}}r_*^{\frac{1}{8}}f^{\frac{3}{5}}$, \citep{fkr02}
where $\dot{M}$ is accretion rate in the units of $10^{17}erg~ s^{-1}$, $m$ is mass of the black hole in units of solar mass, 
 and $f=\bigg(1-\frac{1}{r^\frac{1}{2}}\bigg)^{\frac{1}{4}}$. We use $r=20000$ in general. 
Using Shakura-Sunyaev disk temperature distribution, $T=5\times 10^7K m^{\frac{-1}{2}} \dot{M}^{\frac{1}{4}} r^{\frac{-3}{4}}f^{\frac{1}{4}}$, 
 we find that the temperature at the outer edge is $\sim 10^4K$.
 Here, $\dot{M}$ is the accretion rate of the disk component on the day when advective (halo) component had maximum accretion rate.
\bsp	
\label{lastpage}
\end{document}